# Blockchain and Smart-contracts Modeled in a *SwarmESB* Ecosystem

Ioana Stănescu[1], Lenuța Alboaie[2] and Andrei Panu[3]

[1] Faculty of Computer Science, Alexandru Ioan Cuza University of Iasi, Romania
`stanescu.ioana@info.uaic.ro`
[2] Faculty of Computer Science, Alexandru Ioan Cuza University of Iasi, Romania
`adria@info.uaic.ro`
[3] Faculty of Computer Science, Alexandru Ioan Cuza University of Iasi, Romania
`andrei.panu@info.uaic.ro`

**Abstract.** Blockchain has emerged as a trusted and secure distributed ledger for transactions while also being decentralised, distributed and its legitimacy not guaranteed by a trusted authority. Since the appearance of Bitcoin, Blockchain has known many implementations based on P2P architectures. This paper presents how the blockchain and smart contracts technologies can be integrated into the SwarmESB ecosystem. SwarmESB is a framework that helps building distributed applications, which benefit from privacy and scalability features. Our proposal will present the flexibility in building not only microservices based applications, but also decentralised applications employing blockchain and smart-contracts by modeling a sample Dapp.

## 1 Introduction

In the last years, Blockchain caught a lot of interest because of its initial application as distributed ledger for cryptocurrency (Bitcoin) which proposed a new model for distributed, decentralised applications, which can store information in a secure manner and make it publicly available without risking tampering with the data [1]. Blockchain, together with the concept of smart-contracts [2], opened the path to a new era for web applications, which lead to decentralised applications (Dapps) [3].

Regarding web applications, they have widely transformed in time, evolving from monolithic architectures, to service oriented architectures (SOA), to microservices and serverless. With such a diversity of architectures, which can grow in complexity, a need to ease dealing with concurrency, scalability and integration between components has developed.

The SwarmESB framework was developed in order to provide a method to integrate distributed processes for composing applications while being highly scalable and having privacy embedded in its design [4].

First, we will give an overview of how SwarmESB works and how it can model a distributed application. Then we will present how the blockchain technology can rest on top of the SwarmESB model, how adapters manage the distributed ledger and how smart contracts enhance transactions widening the application's functionality. Lastly, we will present a Dapp sample that depicts the integration between blockchain, smart-contracts and SwarmESB concepts.

## 2      SwarmESB Overview

SwarmESB was developed as a new approach on communication and composability mechanisms for services in cloud systems [5]. It is a framework, which combines the message passing architecture of the Enterprise Service Bus (ESB) with choreographies in order to easily achieve integration between multiple distributed processes. It relies on four basic concepts: messages, swarms (choreographies), adapters and swarm clients [4].

Messages, which are found at the lowest level of the SwarmESB architecture, are sent using pub/sub mechanisms and are responsible with calling the adapters. A group of messages composes the swarms, each message from a swarm is related to a phase of the swarming process which is executed in the context of an adapter. Swarms (choreographies) at functional level are composed of messages. Each step of a choreography can be executed on a different process, without losing the context of the choreography. At structural level, a swarm, which is described in Swarm DSL (language based on javascript), contains four types of constructions: meta (metadata), vars (variables), constructor functions and phases.

Adapters are the long running processes in SwarmESB architecture and represent the component, which provides the actual services. They are responsible with the processing of data or external calls. Adapters having the same functionality can be associated into groups. As it can be seen in Figure 1, between the client and the adapters lays the swarms layer which is responsible with all the logic triggered by a client's request and, through phases, makes calls to adapters.

Swarm clients are the component with which the client application communicates with the swarm based system through WebSockets. The client application can execute a swarm and listen for events sent during the swarm execution. The mechanism is similar with calling an API, which integrates multiple services, the difference being that by swarming privacy can be achieved [7].

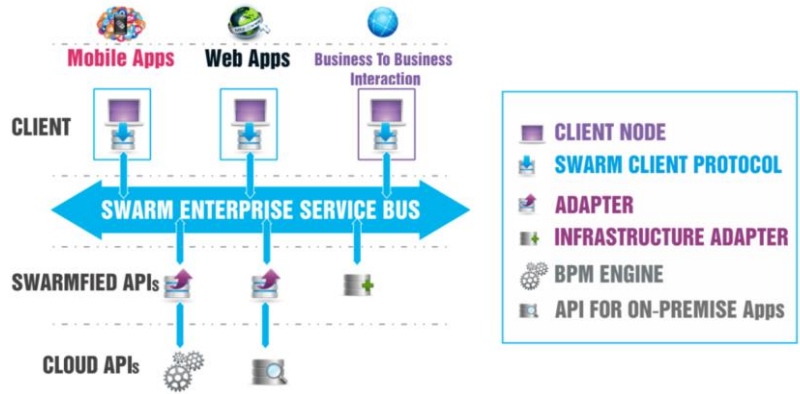

**Fig. 1.** SwarmESB architecture [5]

### 2.1 SwarmESB Enhanced for Blockchain Integration

In order to support building Dapps, a SwarmESB based architecture should at least allow creating P2P like systems and execution of smart-contracts. Considering the SwarmESB architecture, where adapters are the long running processes, which expose services, they would be used as miners and form a P2P network, while swarm choreographies will be used as smart-contracts.

Since adapters cannot directly communicate within a group, the communication will be provided by using swarm choreographies for internal calls in the network. From the point of view of the communication mechanism, swarms need to allow both broadcast and unicast. In the figures below (Figure 2 and Figure 3), a communication flow model is represented using swarm and does a ping-pong communication, both broadcast and unicast. As it can be seen in Figure 2, from the constructor "startBroadcast", the "pingBroadcast" phase is called which executes on every available adapter which sends back "pong" to the calling adapter. Broadcasting in SwarmESB means executing a phase on top of every adapter from within a group (each swarm phase corresponds to a group of adapters), which means that even the calling adapter would receive its own "ping". In the following Figure 3 is an example of direct communication, which executes the "pingDirect" phase on AdapterC who will respond back to AdapterA.

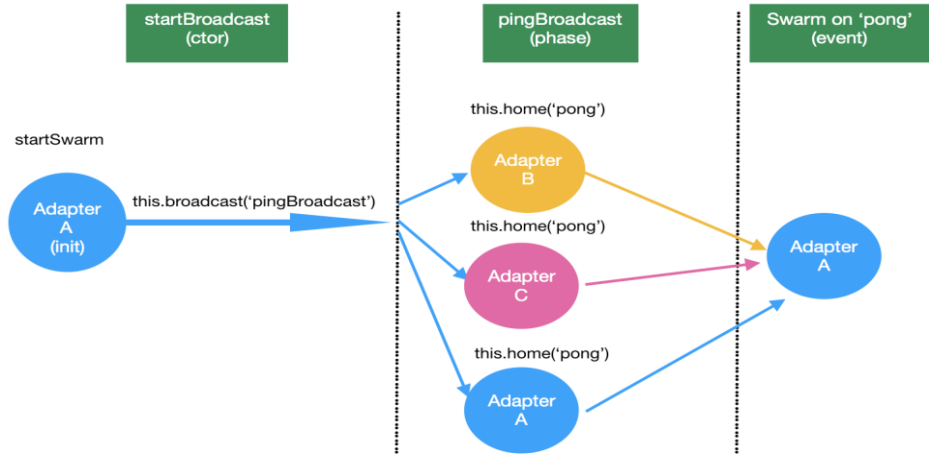

**Fig. 2.** Broadcast in SwarmESB (ping-pong)

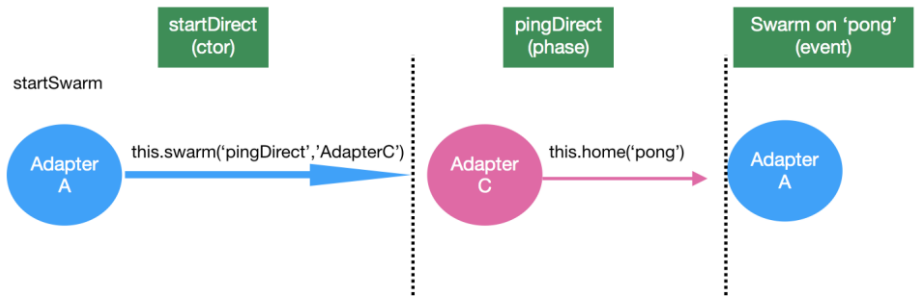

**Fig. 3.** Direct call in SwarmESB (ping-pong)

SwarmESB smart-contracts, as previously mentioned, are implemented using swarm choreographies. A smart-contract, in its basic form, should have a state, be called from the miners and generate transactions called "internal transactions". By using swarms (which are not long running processes), the state cannot be held by the contract, this is an issue that can be solved by keeping a contract's state at miners' level. Another disadvantage is that by not being a long running process swarm contracts cannot issue events or transactions based on a timer or external event, they can only execute based on a transaction.

By providing needed mechanisms for communication and support for smart-contracts, SwarmESB offers a suitable environment for building blockchain based applications. In the following chapter, we will present a detailed model of a decentralised application built on top of SwarmESB.

## 3      Blockchain Application Model using SwarmESB

In our context, the miners are represented by the adapters and the smart-contracts by swarm choreographies. As miners, adapters should support a basic communication protocol composed of:

- broadcast its own identity when joining the network;
- broadcast when a new block is mined;
- require the chain from other peers when local version of the blockchain is unsynced;
- receive transactions from client applications.

The communication will be done through swarms choreographies, and the miners should implement the corresponding methods for the protocol such as: receive newly mined blocks, transactions or peers announcements (register them and respond back), sending current chain and transactions in the pool when requested. The communication scenarios are graphically represented in Figures 4-7.

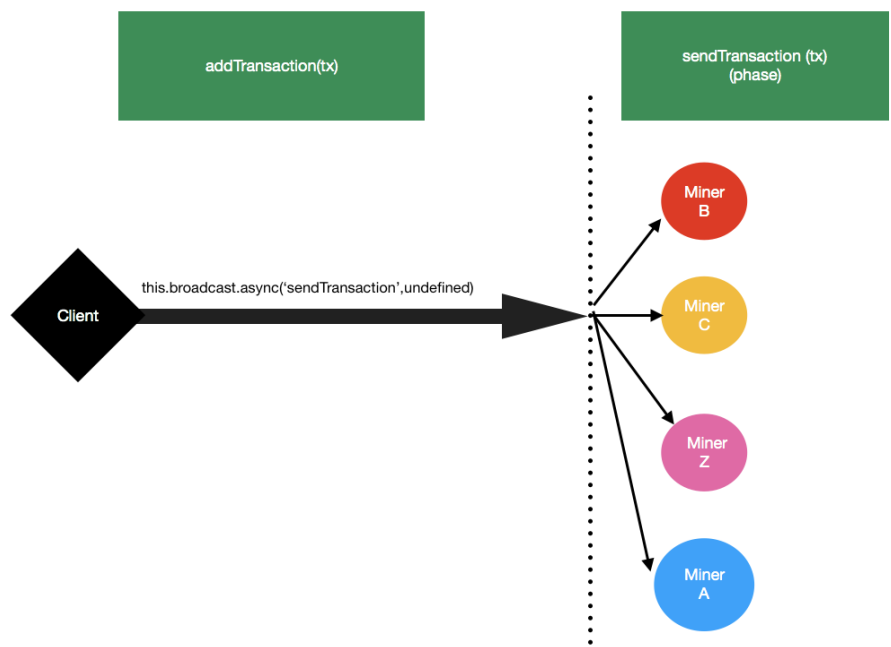

**Fig. 4.** Transaction broadcast from client to miners network

When a new transaction is to be sent to the miners network, the client executes a swarm calling the "addTransaction" constructor and passing the generated transaction. Then, the constructor broadcasts the transaction to the miners via the "sendTransaction" phase, at that point every miner should receive the transaction and add it to their pool of transactions in order to be mined.

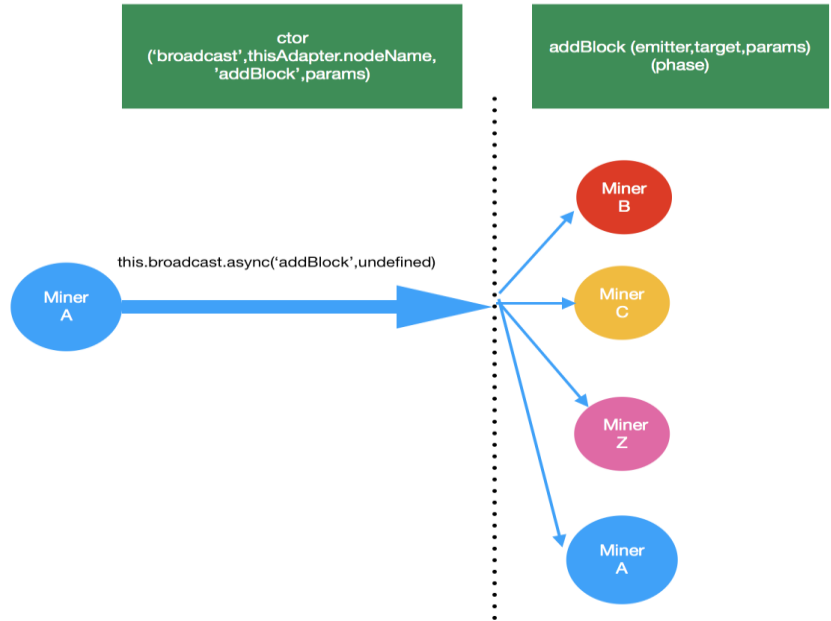

**Fig. 5.** Miner broadcasts mined block

Whenever a new block is mined, the miner will broadcast it to the network by calling the "ctor" with the parameters: "broadcast", its name, "addBlock" (phase name) and the block (params). The constructor then broadcasts a message to the "addBlock" phase in order to make calls to all the other adapters to send them the newly mined block. The miner who emitted the new block will not wait for any responses.

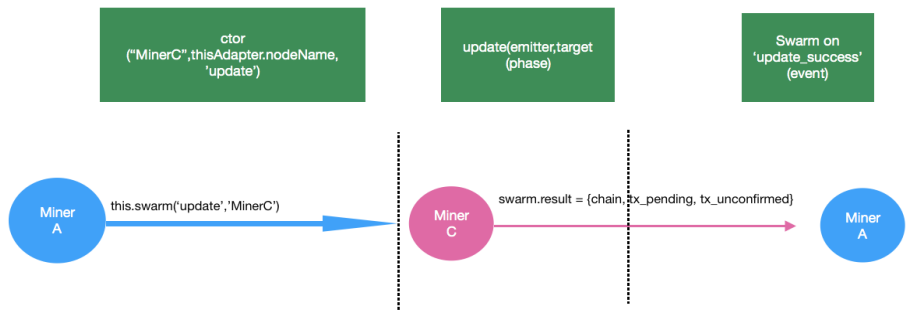

**Fig. 6.** Miner requests update from another miner (direct call)

When a miner has an unsynced chain (it can no longer generate valid blocks for the blockchain), it will call the ctor constructor in order to request the current chain and transactions from a specific miner, to do so it passes the destination's name, its own identity (to identify itself) and "update". From the swarm the update phase will be executed calling the update method from the destination miner, which will return an object composed of the current chain, its pool of pending transactions and the pool of transactions, which await confirmation. The returned result is then added to swarm.result and passed to the calling miner through the update_success event.

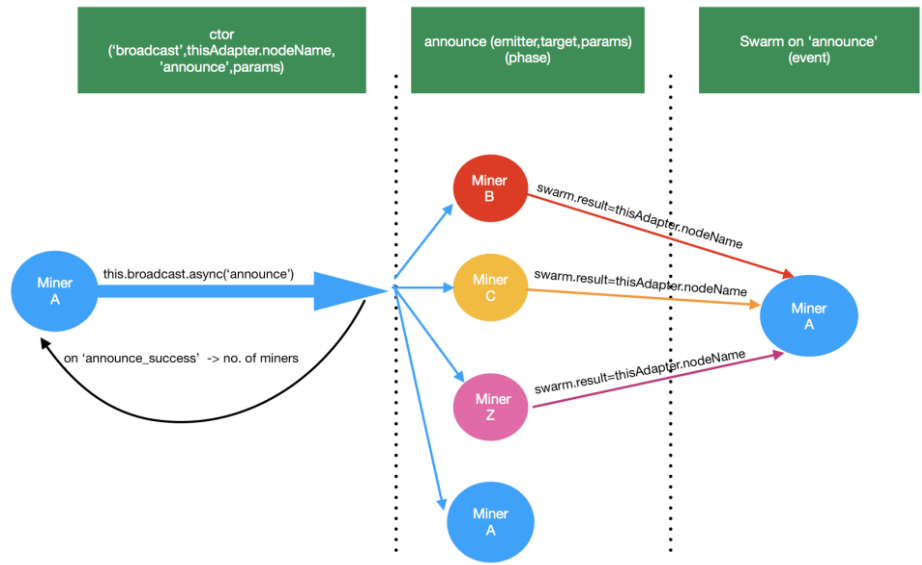

**Fig. 7.** Miner announcing itself in the network

When a new miner joins the network, it will call the constructor "ctor" when executing the swarm choreography used for internal communication, giving as parameter: "broadcast" (the message is meant to be broadcasted to all the peers), its name and "announce" (the name of the phase to be executed) and an optional parameters object containing the chain. From the announce phase, the announce method is called from the miners, after a miner registers the announcement, it will return its own name. The result from the announce method is added to swarm.result, object which is then received by the calling miner when listening for announce event on the executed swarm. When the broadcast call is made, the constructor also sends an "announce_success" event, returning to the calling miner the number of registered peers, so that it will wait for an equal number of "announce" events.

With the communication set in place, mining and validation of transactions, blocks and the chain are trivial and will not be further explained.

### 3.1 Smart-contracts

Regarding smart-contracts, which are a necessary part of a decentralised application based on blockchain, when creating a model, the following facts were taken into consideration: a contract should have (receive) it's previous state, should have access to the chain, receive transaction parameters and should be called from miners(adapters). In the SwarmESB environment, two possible ways to model smart-contracts were available: swarm choreographies and adapters.

Adapters, from the point of view of the functionality (long running processes which can execute tasks based in timers), would be a more appropriate choice, but they would require both instantiating an adapter and creating a swarm choreography which would be used to call methods on adapters. Besides the complexity of the process, adapters as smart-contracts would not be secure, since methods in adapters can be called by any swarm that knows the adapter's name, an adapter's state can be corrupted by a swarm not affiliated with the contract. Because of security and complexity concerns, adapters were not chosen for smart-contracts implementation, instead, smart-contracts would be modeled using swarm choreographies, even though the smart-contracts would not allow for tasks to be executed based on a timer or event, but they would be easily accessible, their state (held by the miner) cannot get corrupted even if that swarm is executed from outside of the network.

Form the point of view of the execution, swarms are called based on their name which means that a contract's address is not calculated by the miners nor the client (based on the transaction which adds the contract in the system), but it actually is the name declared in the metadata of the swarm and the state of the contract will be given by the miner as parameter at execution time (the contract, upon successful execution will return back its current state). Considering these constraints, we propose a model described as follows:

```
var contract =
{
  meta: {
   name: "sampleContract"
  },
  vars: {
   state:null,
   method:null,
   params:null,
   chain:null,
   transaction:null
  },
  ctor: function (transaction,state,chain) {
   this.method = transaction.params.method;
   this.params = transaction.params;
   this.chain = chain;
   this.state = state;
   this.transaction = transaction;
```

```
    this.swarm(this.method); //method is the phase name
    //this.swarm calls the phase given as parameter
   },
   <phase>:{
    node: "All", // node is the adapters group, "All"
means that it can be executed on any adapter
    code: {...} // code to be executed in the current
phase
   }
}
```

The constructor ('ctor') is responsible with instantiating the variables that will be further used in the called phase. In SwarmESB in order to execute a phase, a constructor is called and it is responsible with dispatching the request to the corresponding phase.

As for how a contract is called, the miner will execute the corresponding swarm using a swarm client and will wait for a response. In order not to lock the block mining or validating process, the response from a contract should be awaited for a limited time. In case it does not finish in a timeframe, its result can be set to error or timeout, based on internal logic of miners.

## 4    Conclusions

SwarmESB, from its core architecture, proves itself capable to support not only applications based on microservices, but also decentralised applications based on blockchain and smart-contracts. Even though there are mechanisms that cannot be implemented in the same way as in other blockchain implementations (e.g. smart-contracts are not long running processes and the state it's passed at each swarm initialisation), using SwarmESB also brings its advantages by going one step forward to offer privacy in blockchain applications and by giving more power to smart-contracts, which, through the swarm choreography, can implement complex flows in a less constraining environment.

## Acknowledgements

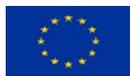The dissemination of this work is partly funded by the European Union's Horizon 2020 research and innovation programme under grant agreement No 692178. It is also partially supported by the Private Sky Project, under the POC-A1-A1.2.3-G-2015 Programme (Grant Agreement no. P 40 371).